\documentclass{jpsj-suppl}
\usepackage{txfonts} 

\usepackage{bm}
\usepackage{sidecap}
\usepackage{color}
\usepackage{ulem}

\renewcommand{\sout}{\bgroup \color[rgb]{1,0,0} \ULdepth=-.5ex \ULset}

\newcommand{\Psfig}[2]{\includegraphics[width=#1]{#2}}
\newcommand{\PsfigII}[2]{\includegraphics[scale=#1]{#2}}

\def\fzero{f_{0}(980)}
\def\azero{a_{0}(980)}
\def\afmix{$\azero$-$\fzero$ }
\def\bmafmix{$\bm{a_{0}(980)}$-$\bm{f_{0}(980)}$ }

\def\kev{\text{ keV}}
\def\mev{\text{ MeV}}

\def\fm{\text{ fm}}

\title{Determining Compositeness of Hadronic Resonances:
  the $\bm{\Lambda (1405)}$ Radiative Decay and the \bmafmix Mixing}

\author{T.~\textsc{Sekihara}$^{1}$ and S.~\textsc{Kumano}$^{2,3}$}

\inst{$^{1}$Research Center for Nuclear Physics
  (RCNP), Osaka University, Ibaraki, Osaka, 567-0047, Japan \\
$^{2}$KEK Theory Center, Institute of Particle and Nuclear
  Studies, High Energy Accelerator Research Organization (KEK), 1-1,
  Oho, Tsukuba, Ibaraki 305-0801, Japan \\
$^{3}$J-PARC Branch, KEK Theory Center,
  Institute of Particle and Nuclear Studies, 
  High Energy Accelerator Research Organization (KEK),
  203-1, Shirakata, Tokai, Ibaraki, 319-1106, Japan}

\email{sekihara@rcnp.osaka-u.ac.jp}

\recdate{October 3, 2014} 

\abst{Recently the concept of compositeness has been developed so as
  to distinguish whether interested hadrons are hadronic molecules or
  not.  Here, in terms of compositeness, we investigate $\bar{K} N$
  molecular structure of the $\Lambda (1405)$ resonance with the
  $\Lambda (1405)$ radiative decay and $K \bar{K}$ molecular structure
  of the $\azero$ and $\fzero$ resonances with the \afmix mixing.}

\kword{hadronic molecules, 
compositeness, 
$\Lambda (1405)$ radiative decay, 
\afmix mixing}

\begin{document}
\maketitle

\section{Introduction}

Although excellent successes of the constituent quark model tell us
that ordinary hadrons consist of three quarks ($qqq$) for baryons and
a quark-antiquark pair ($q \bar{q}$) for mesons~\cite{Olive:1900zz},
there should exist exotic hadrons, which are not able to be classified
as $qqq$ for baryons and $q \bar{q}$ for mesons since the fundamental
theory of strong interaction, QCD, does not prohibit such exotic
systems as long as they are color singlet.  In fact, there are several
experimental indications that some hadrons do not fit into the
classifications by the constituent quark model.  For instance, the
hyperon resonance $\Lambda (1405)$ has an anomalously light mass among
the negative parity baryons and has been expected to be a $\bar{K} N$
molecular state rather than a three-quark state~\cite{Dalitz:1960du}.
Moreover, the lightest scalar meson nonet shows inverted spectrum from
the expectation with the $q \bar{q}$ composition and hence various
exotic configurations have been proposed for the scalar mesons such as
compact tetra-quark states~\cite{Jaffe:1976ig} or $K \bar{K}$
molecules for $\azero$ and $\fzero$~\cite{Weinstein:1982gc}.  In
addition, it is encouraging that charged quarkonium-like states were
observed in the heavy-quark sector by the Belle
collaboration~\cite{Belle:2011aa}.

Among exotic hadrons, hadronic molecules are of special interest since
they are unique in that their constituents are not quarks but hadrons
themselves, which will give various characteristic properties to
hadronic molecules.  For instance, a hadronic molecule can be a
spatially extended object due to the absence of strong quark confining
force.  Actually, spatial size of $\Lambda (1405)$ was theoretically
studied in Refs.~\cite{Sekihara:2008qk, Sekihara:2010uz,
  Sekihara:2012xp} and it was found that its spatial size largely
exceeds the typical hadronic size $\lesssim 0.8 \fm$.  Furthermore,
the uniqueness of hadronic molecules allows us to construct their
two-body wave functions in terms of the hadronic degrees of freedom,
and recently compositeness was introduced as the contribution from the
two-body wave function to the normalization of the total wave function
for hadrons so as to identify hadronic molecules~\cite{Hyodo:2011qc,
  Aceti:2012dd, Hyodo:2013nka, Sekihara:2014kya}.  Since the
compositeness can be evaluated from experimental observables, there is
a possibility to determine experimentally the structure of candidates
of hadronic molecules such as $\Lambda (1405)$.

\section{Dynamically generated hadrons and compositeness}

When a hadron-hadron two-body interaction $V$ is sufficiently strong,
the interaction can dynamically generate a bound state composed of the
two hadrons.  In this condition, the bound state pole appears in the
complex energy plane of the scattering amplitude $T (s)$, which can be
obtained from the Lippmann-Schwinger equation as
\begin{equation}
T ( s ) = V + V G T , 
\label{eq:LSeq}
\end{equation}
with the Mandelstam variable $s$ and the hadron-hadron two-body loop
function $G (s)$.  The bound state pole is described by the pole
position $s_{\rm pole}$ and its residue $g^{2}$ as
\begin{equation}
T ( s ) = \frac{g^{2}}{s - s_{\rm pole}} 
+ (\text{regular at }s = s_{\rm pole}) , 
\label{eq:amp_pole}
\end{equation}
where $g$ can be interpreted as the coupling constant of the bound
state to the hadron-hadron two-body state.  In
Refs.~\cite{Aceti:2012dd, Gamermann:2009uq} it was found that the pole
position $s_{\rm pole}$ and coupling constant $g$ are related to the
two-body wave function.  In the following we treat a two-body
scattering in a relativistic framework, in which the two-body wave
function for the bound state generated with the Lippmann-Schwinger
equation~\eqref{eq:LSeq} is expressed as~\cite{Sekihara:2014kya}
\begin{equation}
\tilde{\Psi} ( \bm{q} ) 
= \frac{g}
{s_{\rm pole} - [\omega (\bm{q}) + \omega ^{\prime}(\bm{q})]^{2}} ,
\quad 
\omega (\bm{q}) \equiv \sqrt{m^{2} + \bm{q}^{2}} ,
\quad 
\omega ^{\prime} (\bm{q}) \equiv \sqrt{m^{\prime 2} + \bm{q}^{2}} ,
\end{equation}
where $m$ and $m^{\prime}$ are masses of the constituent hadrons.  By
using the two-body wave function, we can define compositeness of the
bound state, $X$, as the contribution of the two-body wave function to
the normalization of the total wave function~\cite{Hyodo:2011qc,
  Hyodo:2013nka, Sekihara:2014kya}:
\begin{equation}
X \equiv \int 
\frac{d^{3} q}{( 2 \pi )^{3}}
\frac{\omega (\bm{q}) + \omega ^{\prime}(\bm{q})}
{2 \omega (\bm{q}) \omega ^{\prime}(\bm{q})} 
\left [ \tilde{\Psi} ( \bm{q} ) \right ]^{2}
= - g^{2} \frac{d G}{d s} ( s = s_{\rm pole} ) ,
\label{eq:X}
\end{equation}
where the normalization factor $[\omega _{i}(\bm{q}) + \omega
_{i}^{\prime}(\bm{q})] / [2 \omega _{i}(\bm{q}) \omega
_{i}^{\prime}(\bm{q})]$ guarantees the Lorentz invariance of $X$.  We
note that the compositeness is not an observable and hence is a model
dependent quantity.  On the other hand, components which cannot be
reduced to hadronic two-body configurations, such as compact $q
\bar{q}$ and $q q \bar{q} \bar{q}$ states, contribute to the
elementariness $Z$, and its expression is~\cite{Sekihara:2014kya}
\begin{equation}
Z = - g^{2} \left [ G^{2} \frac{d V}{d s} \right ] _{s = s_{\rm pole}} . 
\label{eq:Z}
\end{equation}
Then the sum of the compositeness $X$ and elementariness $Z$ gives the
normalization of the total wave function as~\cite{Sekihara:2010uz,
  Sekihara:2014kya}
\begin{equation}
  \langle \Psi | \Psi \rangle = X + Z 
  = - g^{2}
  \left [ \frac{d G}{d s} 
    + G^{2} \frac{d V}{d s} 
  \right ] _{s = s_{\rm pole}}
  = 1 .
\label{eq:sum-rule}
\end{equation}
The above discussions can be straightforwardly extended to the
resonance states in coupled-channel approach, but we note that for
resonance states both the compositeness $X$ and elementariness $Z$
becomes complex although the sum is exactly unity, $X + Z = 1$.

Studies on hadron-hadron scatterings with the Lippmann-Schwinger
equation~\eqref{eq:LSeq} and hadronic resonances dynamically generated
as poles in Eq.~\eqref{eq:amp_pole} have been performed especially in
the so-called chiral unitary approach for
meson-meson~\cite{Oller:1997ti} and meson-baryon~\cite{Kaiser:1995eg}
scatterings.  In the chiral unitary approach the interaction $V(s)$ is
taken from chiral perturbation theory and the approach successfully
reproduces the light scalar and vector mesons in the meson-meson
scatterings and $\Lambda (1405)$ in the meson-baryon scatterings.  The
compositeness of these hadronic resonances was theoretically evaluated
in Refs.~\cite{Sekihara:2012xp, Sekihara:2014kya} and it was found that
$\Lambda (1405)$ and $\fzero$ are dominated by the $\bar{K} N$ and $K
\bar{K}$ composite states with the compositeness $X_{\bar{K} N}$ and
$X_{K \bar{K}}$ close to unity, respectively, while compositeness of
other scalar and vector mesons is not large compared to unity.

We here emphasize that, although the compositeness is not an
observable, we can evaluate it from experimental observables via
appropriate models.  In this study we employ the expression in
Eq.~\eqref{eq:X}.  The key is to determine the resonance pole position
of the scattering amplitude and the coupling constant obtained on the
pole position.  Actually the determination can be done by using the
$\Lambda (1405)$ radiative decay for $\Lambda
(1405)$~\cite{Sekihara:2013sma} and the \afmix mixing for $\azero$ and
$\fzero$~\cite{Sekihara:2014qxa}, as we will discuss below.

\section{The $\bm{\bar{K} N}$ compositeness of  $\bm{\Lambda (1405)}$
  from its radiative decay width}

Although the hyperon resonance $\Lambda (1405)$ decays to $\pi \Sigma$
with the branching ratio $100 \%$~\cite{Olive:1900zz}, the radiative
decay of $\Lambda (1405)$, $\Lambda (1405) \to \Lambda \gamma$ and
$\Sigma ^{0} \gamma$, is in principle possible.  Indeed, in PDG there
are ``experimental'' data on the $\Lambda (1405)$ radiative decay
width evaluated from an isobar model fitting of the decays of the
$K^{-} p$ atom~\cite{Burkhardt:1991ms}: $\Gamma _{\Lambda \gamma} = 27
\pm 8 \kev$ and $\Gamma _{\Sigma ^{0} \gamma} = 10 \pm 4 \kev$ or $23
\pm 7 \kev$, which imply that the branching ratio of the radiative
decay is $\sim 0.1 \%$.  The $\Lambda (1405)$ radiative decay is
closely related to the structure of $\Lambda (1405)$ as an $E1$
transition, and here we investigate a relation between the $\bar{K} N$
compositeness for $\Lambda (1405)$ and its radiative decay width via
the $\Lambda (1405)$-$\bar{K} N$ coupling constant as a free
parameter.

\begin{figure}[t]
  \centering
  \begin{tabular}{ccc}
    \PsfigII{0.18}{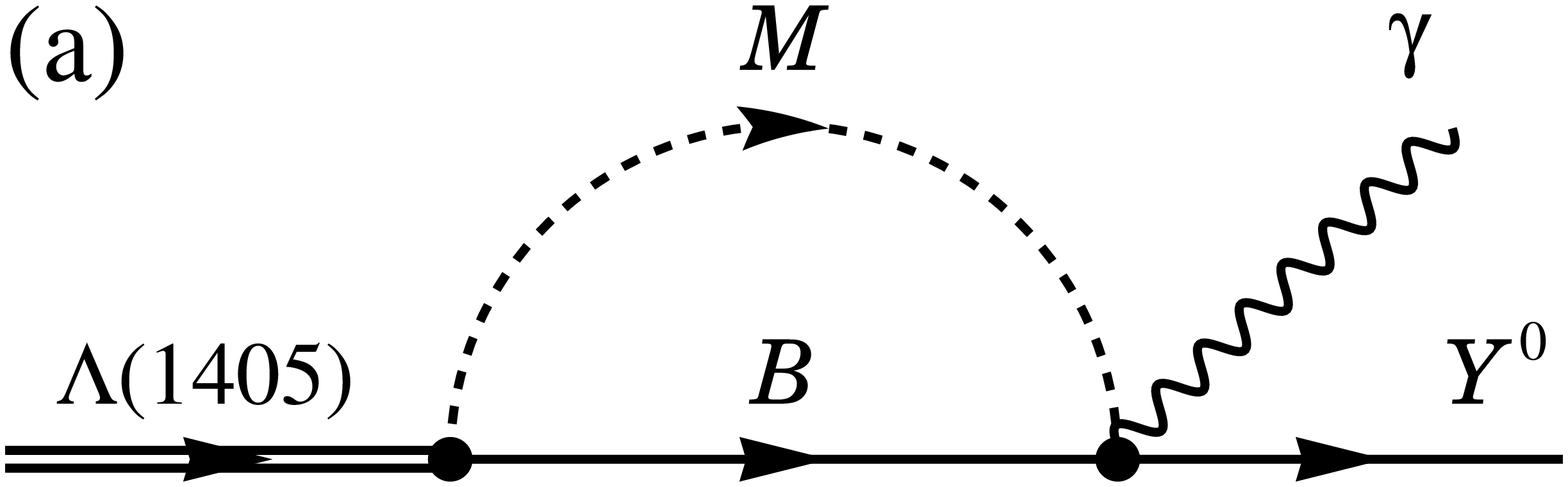} & 
    \PsfigII{0.18}{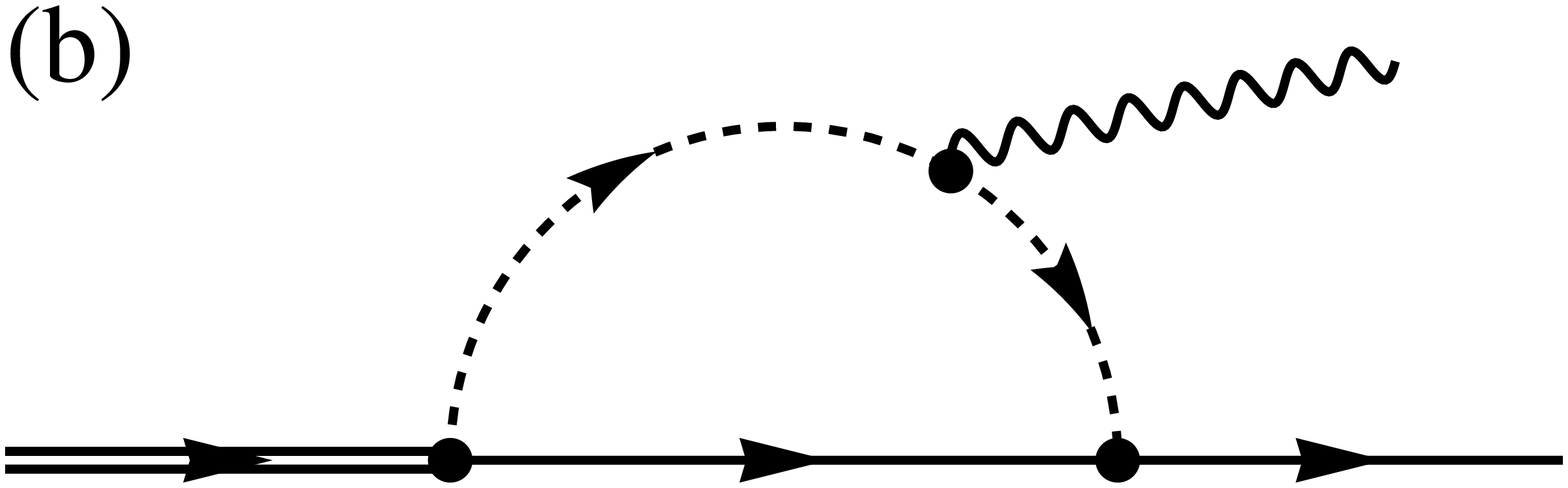} &
    \PsfigII{0.18}{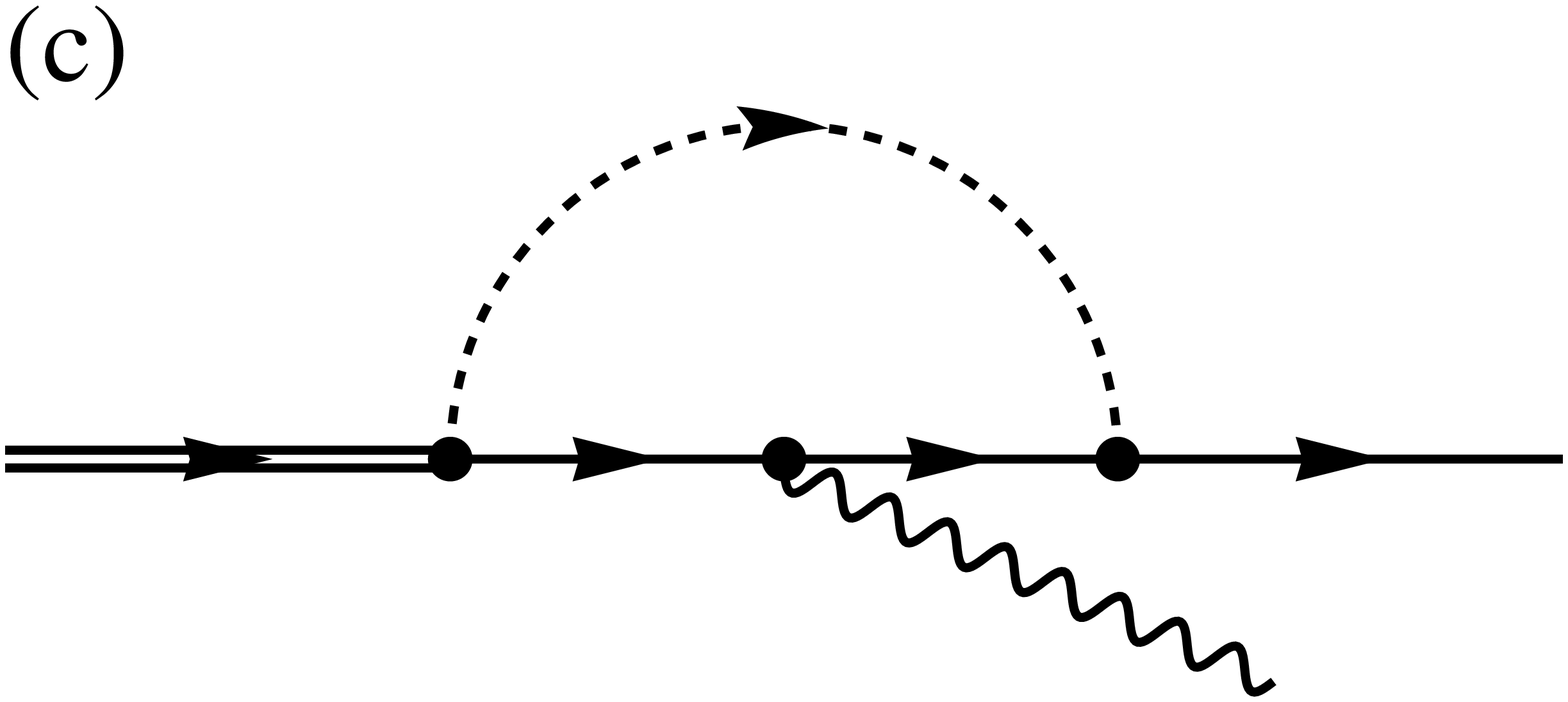} 
  \end{tabular}
  \caption{Feynman diagrams for the $\Lambda (1405)$ radiative decay
    employed in this study~\cite{Sekihara:2013sma}.  In the figure (a)
    $M$ and $B$ denote mesons and baryons, respectively, and $Y^{0}$
    is $\Lambda$ or $\Sigma ^{0}$. }
  \label{fig:1}
  \vspace{-15pt}
\end{figure}

Our formulation of the $\Lambda (1405)$ radiative decay is based on
that developed in Ref.~\cite{Geng:2007hz}.  The Feynman diagrams
relevant to the radiative decay are shown in Fig.~\ref{fig:1}, which
is obtained in the picture that the photon is emitted from
meson-baryon components inside $\Lambda (1405)$ without considering
compact $q q q$ nor $q q q q \bar{q}$ state for $\Lambda (1405)$.  As
one can see from the diagrams, the decay amplitude contains coupling
constants of $\Lambda (1405)$ to meson-baryon states, thus we can
relate the compositeness of $\Lambda (1405)$ with the decay width via
the coupling constant [see Eq.~\eqref{eq:X}].  In this study we fix
the $\Lambda (1405)$ pole position by using the mass and width taken
from PDG~\cite{Olive:1900zz}: $s_{\rm pole} = ( M_{\Lambda (1405)} - i
\Gamma _{\Lambda (1405)}/2 )^{2}$ with $M_{\Lambda (1405)} = 1405
\mev$ and $\Gamma _{\Lambda (1405)} = 50 \mev$.  The absolute value of
the $\Lambda (1405)$-$\bar{K} N$ coupling constant is determined by
the absolute value of the $\bar{K} N$ compositeness with
Eq.~\eqref{eq:X}, while the $\Lambda (1405)$-$\pi \Sigma$ coupling
constant is fixed by the $\Lambda (1405)$ decay width, $\Gamma
_{\Lambda (1405)} = 50 \mev$.  The relative phase between $\Lambda
(1405)$-$\bar{K} N$ and $\pi \Sigma$ coupling constants is not known,
hence we calculate both maximally constructive and destructive
conditions in order to evaluate the allowed range of the radiative
decay width.
We finally note that the divergence coming from each diagram in
Fig.~\ref{fig:1} is cancelled when contributions from three diagrams
are summed due to the gauge and Lorentz symmetries.

\begin{figure}[t]
  \vspace{-15pt}
  \centering
  \begin{minipage}{0.5\hsize}
    \Psfig{7.2cm}{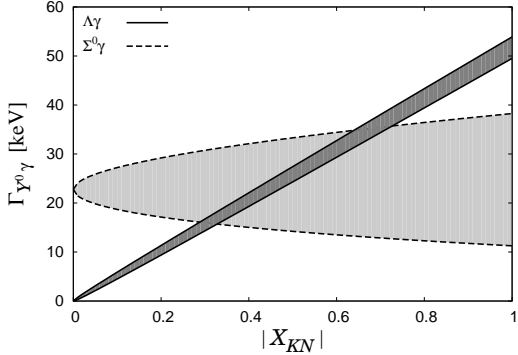}
  \end{minipage}
  \begin{minipage}{0.45\hsize}
    \caption{Allowed range of the $\Lambda (1405)$ radiative decay
      width $\Gamma _{Y^{0} \gamma}$ with respect to the $\bar{K} N$
      compositeness $| X_{\bar{K} N} |$~\cite{Sekihara:2013sma}.}
  \end{minipage}
  \vspace{-15pt}
\end{figure}

In this strategy we can calculate the allowed range of the $\Lambda
(1405)$ radiative decay width with respect to each value of the
absolute value of the $\bar{K} N$ compositeness, and the result is
shown in Fig.~2. 
The most prominent property of the radiative decay width can be seen
in the $\Lambda \gamma$ decay mode.  Namely, the allowed range of the
$\Lambda \gamma$ decay width increases almost linearly with a small
band as the absolute value of the $\bar{K} N$ compositeness,
$|X_{\bar{K}N}|$, increases.  This behavior comes from the fact that
the contributions from the $\pi ^{+} \Sigma ^{-}$ and $\pi ^{-} \Sigma
^{+}$ components are largely cancelled with each other in the $\Lambda
\gamma$ decay mode and hence the $K^{-} p$ component dominates the
$\Lambda \gamma$ decay, as discussed in Ref.~\cite{Geng:2007hz}.
Therefore, a large $\Lambda \gamma$ decay width implies a large
$\bar{K} N$ component inside $\Lambda (1405)$ and thus the radiative
decay $\Lambda (1405) \to \Lambda \gamma$ is suited to investigate the
$\bar{K} N$ component.  On the other hand, the $\Sigma ^{0} \gamma$
decay width is nonzero even when the $\bar{K} N$ component is absent,
$|X_{\bar{K} N}|=0$, since there is no cancellation between
contributions from the $\pi ^{\pm} \Sigma ^{\mp}$ component in
contrast to the $\Lambda \gamma$ decay mode.  Then comparing with the
``experimental'' value of the decay width, we can estimate the
$\bar{K} N$ compositeness as $| X_{\bar{K} N} | = 0.5 \pm 0.2$ from
$\Gamma _{\Lambda \gamma} = 27 \pm 8 \kev$, $| X_{\bar{K} N} | > 0.5$
from $\Gamma _{\Sigma ^{0} \gamma} = 10 \pm 4 \kev$, and $| X_{\bar{K}
  N} |$ can be arbitrary for $\Gamma _{\Sigma ^{0} \gamma} = 23 \pm 7
\kev$. This estimation implies that $\bar{K} N$ seems to be the
largest component inside $\Lambda (1405)$.  We note that the $\Lambda
(1405)$ pole position dependence of our result is small as discussed
in Ref.~\cite{Sekihara:2013sma}.

In this study we have neglected compact $q q q$ and $q q q q \bar{q}$
configurations for $\Lambda (1405)$, by assuming that $\Lambda (1405)$
is dominated by the meson-baryon components.  However, in general we
have to take into account them when we calculate the radiative decay
width with $| X_{\bar{K} N} | <1$.  Thus, here we roughly estimate the
behavior of the $\Lambda \gamma$ decay width taking into account the
$q q q$ configuration for $\Lambda (1405)$ as well.  On the one hand,
as we have seen above, hadronic molecules can have a radiative decay
width of $\lesssim 100 \kev$.  On the other hand, excited baryons
which are expected to have a $q q q$ configuration, such as $N
(1535)$, have radiative decay widths of $\sim 100 \kev$--$1 \mev$,
which implies that the radiative decay width accompanied by a quark
transition inside the hadron would be relatively large.  As a
consequence, when we describe $\Lambda (1405)$ as a mixture of the
compact $q q q$ state and the $\bar{K} N$ molecular state with the
correct normalization of the total wave function~\eqref{eq:sum-rule},
the slope of $\Gamma _{\Lambda \gamma}$ would be negative with respect
to $| X_{\bar{K} N} |$ and the decay width becomes, for instance,
$\Gamma _{\Lambda \gamma} \sim 200 \kev$ for $| X_{\bar{K} N} | = 0$
and $\Gamma _{\Lambda \gamma} \sim 50 \kev$ for $| X_{\bar{K} N} | =
1$.  In this sense, the small ``experimental'' decay width $\Gamma
_{\Lambda \gamma} = 27 \pm 8 \kev$ for $\Lambda (1405)$ might be a
signal that $\Lambda (1405)$ is a hadronic molecule.

\section{The $\bm{K \bar{K}}$ compositeness of $\bm{\azero}$ and
  $\bm{\fzero}$ from their mixing intensity}

\begin{figure}[t]
  \centering
  \begin{tabular}{ccc}
    \PsfigII{0.18}{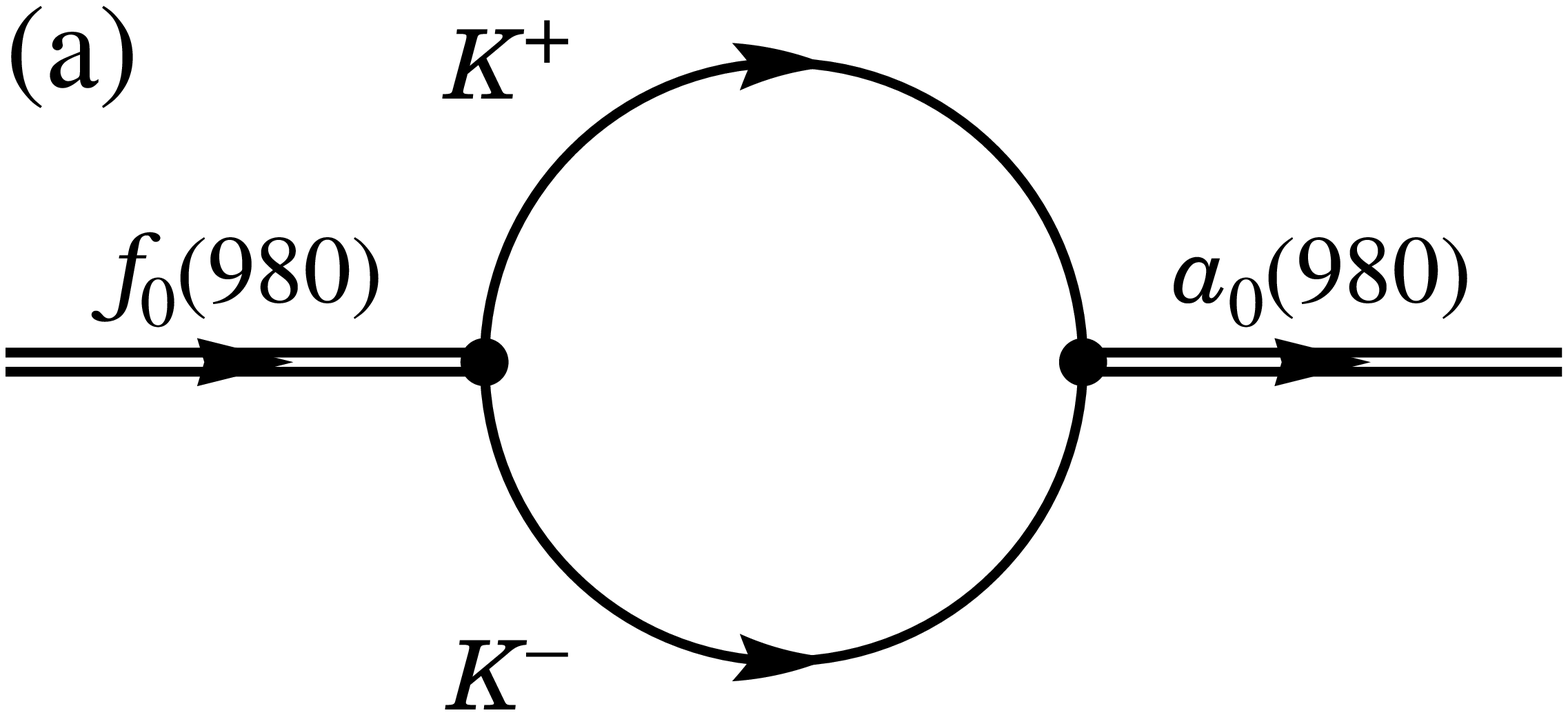} & 
    \PsfigII{0.18}{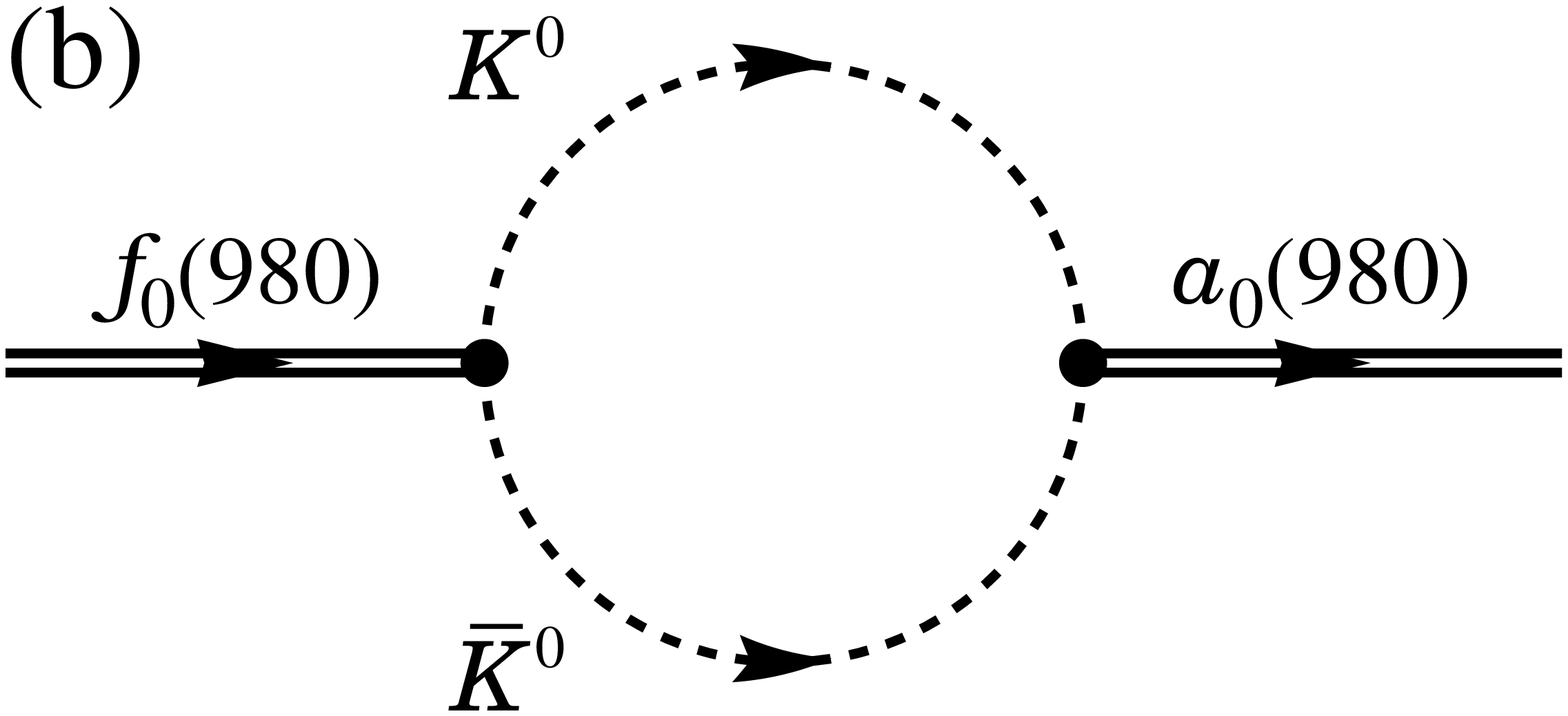} &
    \PsfigII{0.18}{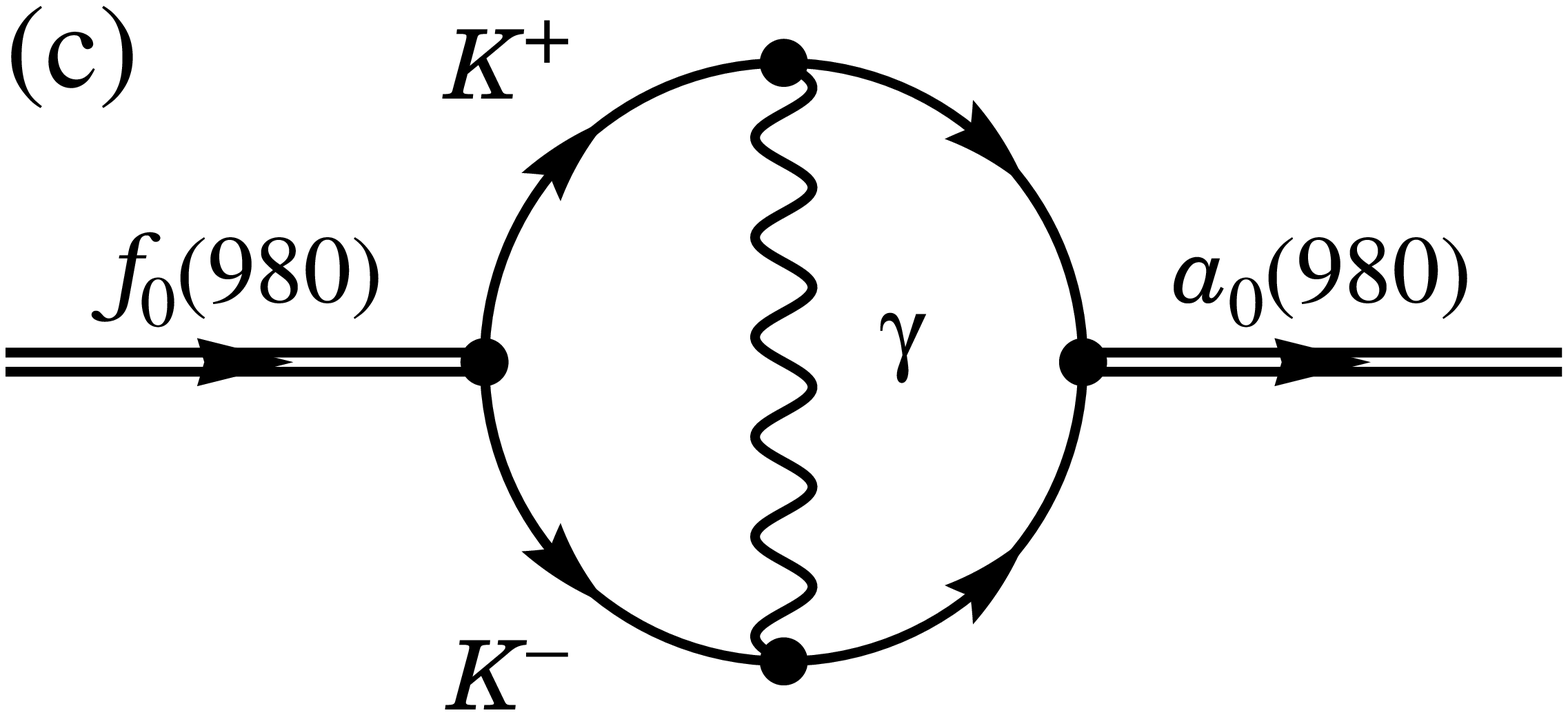} 
  \end{tabular}
  \caption{Feynman diagrams for the \afmix mixing employed in this
    study~\cite{Sekihara:2014qxa}.}
  \label{fig:3}
  \vspace{-15pt}
\end{figure}

Since two scalar mesons $\azero$ and $\fzero$ have almost degenerate
masses, there is a possibility that $\azero$ and $\fzero$ are mixed
with each other in processes which break isospin symmetry.  Especially
it was pointed out in Ref.~\cite{Achasov:1979xc} that the mixing
effect should be unusually enhanced around the $K \bar{K}$ threshold
due to the difference of the phase spaces of $K^{+} K^{-}$ and $K^{0}
\bar{K}^{0}$ [see Fig.~\ref{fig:3}(a) and (b)].  The \afmix mixing
effect was recently observed in Ref.~\cite{Ablikim:2010aa}, in which
decay of $J / \psi$ to $\phi \pi \eta$ was used to confirm the mixing
effect.  In Ref.~\cite{Ablikim:2010aa} the mixing intensity $\xi _{f
  a}$ was introduced as
\begin{equation}
  \xi _{f a} \equiv 
  \frac{\displaystyle \text{Br} ( J/\psi \to \phi f_{0}(980) 
    \to \phi a_{0}^{0}(980) \to \phi \pi ^{0} \eta )}
  {\displaystyle \text{Br} ( J/\psi \to \phi f_{0}(980) \to \phi \pi \pi )} , 
\label{eq:xi_fa}
\end{equation}
and its value was obtained as $\xi _{f a} = 0.60 \pm 0.20
(\text{stat}) \pm 0.12 (\text{sys}) \pm 0.26 (\text{para}) \%$ and
$\xi _{f a} |_{\text{upper limit}} = 1.1 \% \text{ (90\%
  C.L.)}$~\cite{Ablikim:2010aa}.

The \afmix mixing should be closely related to the $K \bar{K}$
structure of the scalar mesons $\azero$ and $\fzero$ since the mixing
amplitude contains the $\azero$-$K \bar{K}$ and $\fzero$-$K \bar{K}$
coupling constants.  Therefore, through the $\azero$- and $\fzero$-$K
\bar{K}$ coupling constants we can investigate the $K \bar{K}$
component inside the scalar mesons $\azero$ and $\fzero$ from their
mixing intensity $\xi _{f a}$.  Here we give a constraint on the $K
\bar{K}$ compositeness of $\azero$ and $\fzero$, which we denote as
$X_{a}$ and $X_{f}$, respectively, from the \afmix mixing intensity.

For the calculation of the \afmix mixing we employ three Feynman
diagrams shown in Fig.~\ref{fig:3}.  The diagrams (a) plus (b) in
Fig.~\ref{fig:3} contribute to the mixing at the leading order, while
the diagram (c) in Fig.~\ref{fig:3} gives a subleading order
contribution.  From these diagrams we nonperturbatively calculate the
full propagators of $\azero$ and $\fzero$ and the transition between
$\azero$ and $\fzero$ by summing up all the contributions $\azero \to
\fzero \to \azero \to \cdots$.  The mixing intensity $\xi _{f a}$ is
calculated as the ratio of the two partial decay widths according to
the experimental analysis in Eq.~\eqref{eq:xi_fa}.  In this
formulation the model parameters for the mixing intensity $\xi _{f a}$
are the masses and widths of the scalar mesons, [$M_{a}$ and $\Gamma
_{a}$ for $\azero$ and $M_{f}$ and $\Gamma _{f}$ for $\fzero$,
respectively], and the $\azero$- and $\fzero$-$K \bar{K}$ coupling
constants.  In our strategy, we appropriately fix the masses and
widths of the scalar mesons, while the values of the $\azero$- and
$\fzero$-$K \bar{K}$ coupling constants move independently within
appropriate ranges so as to evaluate simultaneously the mixing
intensity $\xi _{f a}$ and the absolute value of the $K \bar{K}$
compositeness of $\azero$ and $\fzero$, $|X_{a}|$ and $|X_{f}|$,
respectively, with Eq.~\eqref{eq:X}.

\begin{figure}[b]
  \centering
  \begin{minipage}{0.5\hsize}
    \Psfig{7.2cm}{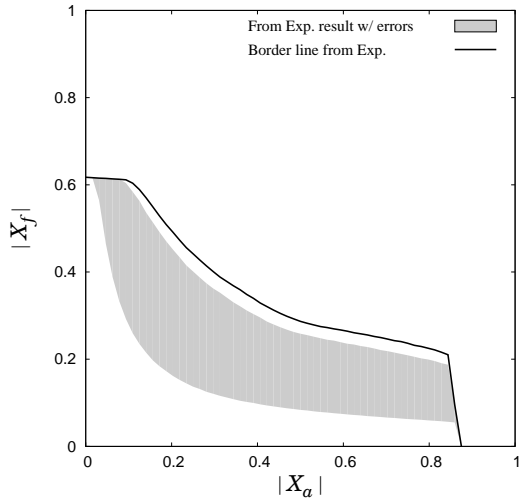} 
  \end{minipage}
  \begin{minipage}{0.45\hsize}
    \caption{Allowed region for the absolute value of compositeness,
      $|X_{a}|$ and $|X_{f}|$, constrained by the experimental value
      of the \afmix mixing intensity~\cite{Sekihara:2014qxa}.  The
      shaded area corresponds to the allowed region by the
      experimental mixing intensity with errors, $\xi _{f a} = 0.60
      \pm 0.20 \pm 0.12 \pm 0.26 \%$.  The solid line indicates the
      border line for the mixing intensity $\xi _{f a} = 1.1 \%$.}
  \end{minipage}
  \label{fig:4}
\end{figure}

Now we fix the parameters $M_{a} = M_{f} = 980 \mev$, $\Gamma _{a} =
100 \mev$, and $\Gamma _{f} = 50 \mev$, calculate both the mixing
intensity and the $K \bar{K}$ compositeness, and investigate the
values of $|X_{a}|$ and $|X_{f}|$ favored by the experimental mixing
intensity.  The result of the allowed region for $|X_{a}|$ and
$|X_{f}|$ in the $|X_{a}|$-$|X_{f}|$ plane is plotted in
Fig.~4.  In the figure the shaded area corresponds to the
allowed region by the experimental mixing intensity $\xi _{f a} = 0.60
\pm 0.20 \pm 0.12 \pm 0.26 \%$, and the solid line to the border line
for the mixing intensity $\xi _{f a} |_{\text{upper limit}} = 1.1 \%$.
As one can see from the figure, the region $|X_{a}| \approx |X_{f}|
\approx 1$ is not favored, which excludes the possibility that both
$\azero$ and $\fzero$ are simultaneously $K \bar{K}$ molecules.  We
have also examined several parameter sets and found that in most
parameter sets $|X_{a}|$ and $|X_{f}|$ cannot be simultaneously take
values close to unity with the constraint of the experimental mixing
intensity~\cite{Sekihara:2014qxa}.  This implies that the statement
that both $\azero$ and $\fzero$ are simultaneously $K \bar{K}$
molecules is questionable.

\section{Summary}

In this study we have investigated hadronic molecular structure of
$\Lambda (1405)$, $\azero$, and $\fzero$ in terms of compositeness,
which is defined as the contribution of the two-body wave function to
the normalization of the total wave function and measures amount of
the two-body component inside the hadron.  We have seen that the
compositeness is not only a theoretical concept but also a quantity
which can be evaluated from experimental observables via appropriate
models.  For instance, from the $\Lambda (1405)$ radiative decay width
we have obtained an implication that the $\bar{K} N$ compositeness of
$\Lambda (1405)$ is $| X_{\bar{K} N} | \gtrsim 0.5$, and from the
\afmix mixing intensity we have found that the statement that both
$\azero$ and $\fzero$ are simultaneously $K \bar{K}$ molecules is
questionable.

At last we emphasize again that we can investigate hadronic molecular
structure without relying directly upon the underlying theory, QCD,
since the constituents of hadronic molecules are hadrons themselves,
which are color-singlet states and hence observable.  Due to this
fact, one can in principle construct hadronic two-body wave functions
with hadronic degrees of freedom and use the wave functions to
evaluate the two-body component inside the hadronic molecules.  In
this sense, the $\Lambda (1405)$ radiative decay and the \afmix mixing
are exactly the experimental phenomena with which one can discuss the
structure of hadronic resonances without considering QCD directly.

T.~S. acknowledges the organizers of J-PARC 2014 symposium for the
financial support to his visit to the conference venue.
S.~K. was supported by the MEXT KAKENHI Grant No. 25105010.

\end{document}